\documentclass{article}

\usepackage{microtype}
\usepackage{graphicx}
\usepackage{subfigure}
\usepackage{booktabs} 
\usepackage{bm}
\usepackage[accepted,nohyperref]{icml2019}

\icmltitlerunning{Image Segmentation of Liver Stage Malaria Infection with Spatial Uncertainty Sampling}

\begin{document}

\twocolumn[
\icmltitle{Image Segmentation of Liver Stage Malaria Infection \\ with Spatial Uncertainty Sampling
}




\begin{icmlauthorlist}
\icmlauthor{Ava P. Soleimany}{mit,har}
\icmlauthor{Harini Suresh}{mit}
\icmlauthor{Jose Javier Gonzalez Ortiz}{mit}
\icmlauthor{Divya Shanmugam}{mit}
\icmlauthor{Nil Gural}{mit}
\icmlauthor{John Guttag}{mit}
\icmlauthor{Sangeeta N. Bhatia}{mit,hhmi}
\end{icmlauthorlist}

\icmlaffiliation{mit}{Massachusetts Institute of Technology, Cambridge, Massachusetts, USA}
\icmlaffiliation{har}{Harvard University, Cambridge, Massachusetts, USA}
\icmlaffiliation{hhmi}{Howard Hughes Medical Institute, Cambridge, Massachusetts, USA}

\icmlcorrespondingauthor{Ava P. Soleimany}{asolei@mit.edu}
\icmlkeywords{image segmentation, model uncertainty, liver stage malaria}

\vskip 0.3in
]



\printAffiliationsAndNotice{}  

\begin{abstract}
Global eradication of malaria depends on the development of drugs effective against the silent, yet obligate liver stage of the disease.
The gold standard in drug development remains microscopic imaging of liver stage parasites in \textit{in vitro} cell culture models. Image analysis presents a major bottleneck in this pipeline since the parasite has significant variability in size, shape, and density in these models. 
As with other highly variable datasets, traditional segmentation models have poor generalizability as they rely on hand-crafted features; thus, manual annotation of liver stage malaria images remains standard.
To address this need, we develop a convolutional neural network architecture that utilizes spatial dropout sampling for parasite segmentation and epistemic uncertainty estimation in images of liver stage malaria. 
Our pipeline produces high-precision segmentations nearly identical to expert annotations, generalizes well on a diverse dataset of liver stage malaria parasites, and promotes independence between learned feature maps to model the uncertainty of generated predictions. 
\end{abstract}

\section{Introduction}
\label{introduction}

Malaria remains a major global health scourge, with nearly half of the world's population remaining at risk \cite{world2018world}. \textit{Plasmodium vivax} is the main barrier to malaria eradication because it harbors dormant forms in the liver, termed hypnozoites, which can reactivate weeks to years after the initial infection and cause relapsing disease. Thus, a malaria eradication campaign cannot be envisioned without eliminating the hypnozoite reservoir. However, there is only one clinically available drug that has anti-hypnozoite activity, which underlies the pressing need for continued drug screening and development \cite{wells2010targeting,alonso2011research}.

Current drug development approaches for liver stage anti-malarials involve large-scale screening against \textit{in vitro} cultures of liver cells that have been infected with malaria parasites \cite{gural2018vitro, antonova2018open, meister2011imaging, roth2018comprehensive}, with immunofluorescent microscopy as a primary readout. Though manual image segmentation for parasite identification and sizing remains standard, it is time and labor intensive, requiring significant technical expertise. 

To address this need for automation in the anti-malarial screening pipeline, we present a convolutional neural network-based (CNN) architecture for automated segmentation of parasites in liver stage malaria infection, and develop a Bayesian deep learning approach for estimating uncertainty in image segmentations. We demonstrate accurate parasite detection and segmentation on a challenging liver stage \textit{P. vivax} dataset, characterized by its variability in terms of parasite shape, density, and size. Finally, through dropout sampling of feature maps during training, we estimate the uncertainty of our segmentations and develop a generalizable algorithm for epistemic, or model, uncertainty~\cite{kendall2017uncertainties} metrics in image segmentation.

\begin{figure}[!h]
\includegraphics[width=\linewidth]{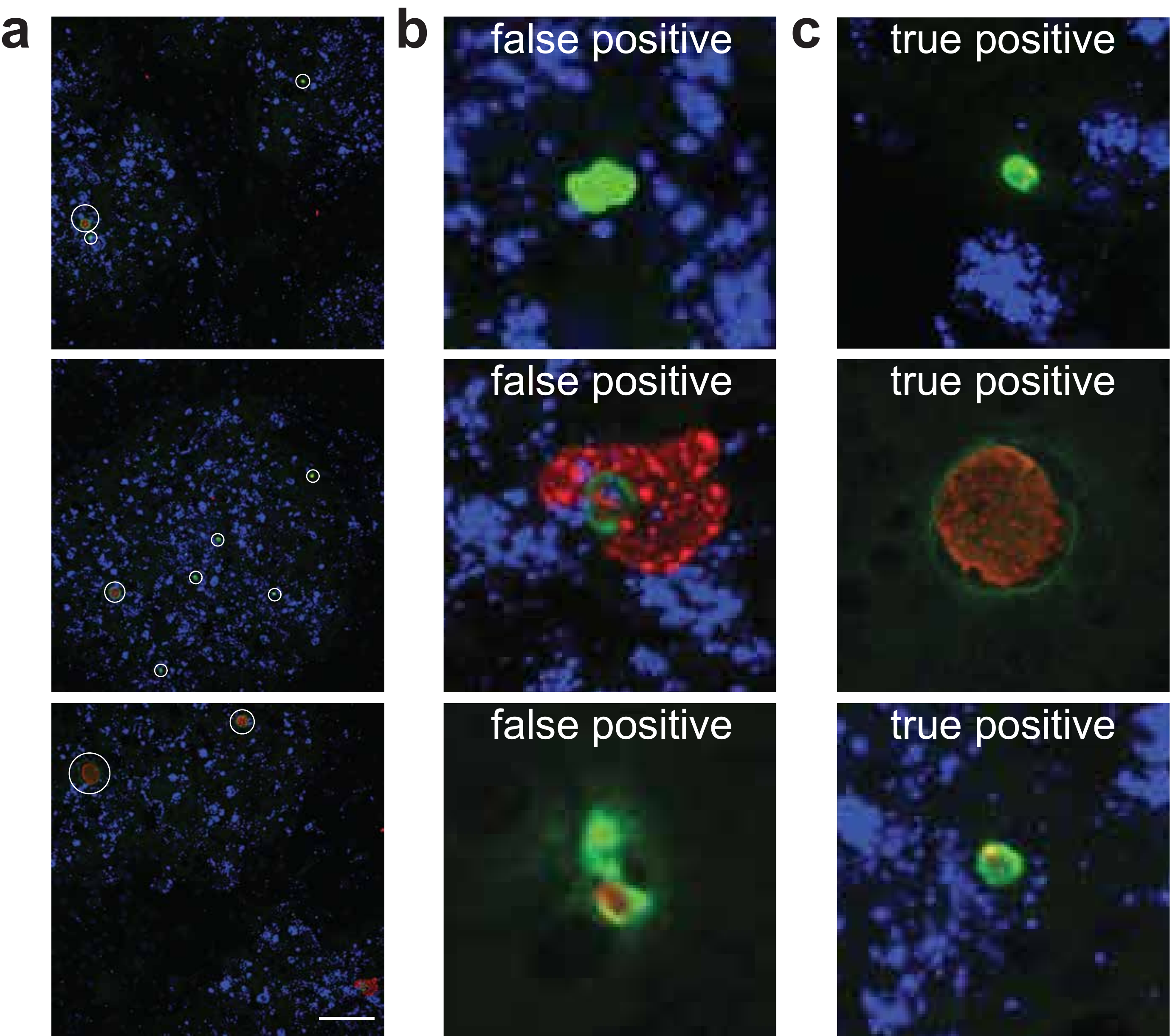}
\caption{{\bf Dataset of liver stage \textit{Plasmodium vivax} infection.}
Representative images from dataset of \textit{P. vivax} infected human hepatocyte cultures, eight days post infection. Green and red channels correspond to stains for parasite membrane and cytoplasmic proteins, respectively, and blue channel corresponds to nuclear stain. A: Representative data, where white circles indicate true positives. Scale bar is 100 $\mu$m. B: Representative examples of false positives in image patches. C: Representative examples of true positives in image patches, including the small forms or hypnozoites (top \& bottom) and large forms or schizonts (middle).}
\vskip -0.2in
\label{dataset}
\end{figure}

\section{Related Work}
\label{relatedwork}
Approaches to biomedical image segmentation fall into rule-based and network-based methods. A common rule-based approach is thresholding based on the intensity of fluorescent stains, which suffers in the face of noise and non-uniform stain intensities in the images \cite{phansalkar2011adaptive}. A popular open-source software, CellProfiler uses user-defined features in conjunction with image processing algorithms to perform object identification for cell image analysis \cite{carpenter2006cellprofiler}. CellProfiler can achieve high accuracies on many image processing problems, yet is limited in its generalizability and robustness, as a new pipeline and a new feature set must be created for each distinct task.

CNNs have emerged as an attractive approach for the segmentation of biomedical images due to their strength on visual recognition tasks. For example, a convolutional network with a sliding-window setup to predict the class label of each pixel in an image has been presented \cite{ciresan2012deep}, yet this approach suffers from computational inefficiency and requires significant spatial context to achieve high accuracy. Our approach addresses these limitations by using a convolutional U-Net \cite{ronneberger2015u} for accurate, efficient image analysis of liver stage malaria infection. Furthermore, we develop a generalizable method for spatial uncertainty modeling \cite{amini2018spatial} in image segmentation and provide concrete estimates of model uncertainty on our malaria dataset.

\section{Methodology}

Given a dataset of input image and binary segmentation pairs $(\bm X, \bm Y)$, where $\bm Y$ is a pixel-wise binary classification of the input image with positive pixel labels corresponding to the parasite class, we learn a functional mapping $\bm f$ parameterized by weights $\bm W$ such that the distance from its output $\bm {\hat Y}=\bm f(\bm X;\bm W)$ to the true labels $\bm Y$ is minimized. 

\subsection{Dataset}
The liver stage \textit{P. vivax} dataset consists of images of human hepatocytes cultured in the micropatterned co-culture format and imaged eight days following infection with \textit{P. vivax} parasites. Cultures were fixed and stained for the parasite membrane protein UIS4 (green channel) and either a histone acetylation marker, H3K9ac, or a cytoplasmic parasite protein, BIP (red channel). DAPI (blue channel) staining was used to mark hepatocyte nuclei. On day eight, cultures comprise two parasite stages: the dormant hypnozoites which are small in size and uninucleated (Fig.~\ref{dataset}c, top \& bottom), and the maturing schizonts which are large and multinucleated (Fig.~\ref{dataset}c, middle). Approximately 0.2-1\% of hepatocytes are infected based on the typical infectivity of \textit{P. vivax} in this culture system. All images were annotated by experienced researchers via manual identification and segmentation of the parasites present. The dataset was split into training, validation, and test cohorts, and $256 \times 256$ pixel RGB image patches were selected from each $1024\times1024$ images. Patches that contained the presence of the green UIS4 parasite membrane stain and the red BIP parasite protein stain above a threshold $T$ were fed into the model for training. We further augmented the data by applying random rotations and symmetries. Approximately $5000$ augmented image patches were used for training. 

\subsection{Model}
\begin{figure*}[!h]
\includegraphics[width=\linewidth]{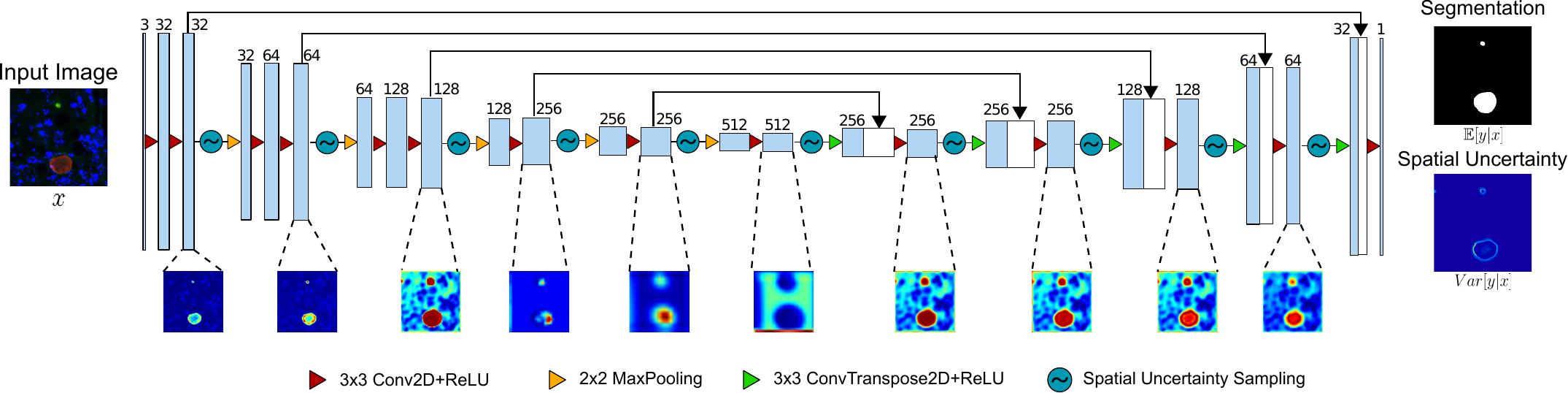}
\caption{{\bf Architecture for segmentation of images of liver stage malaria infection.} A convolutional U-Net successively compresses and decompresses the input image to output a pixel-wise classification. Convolutional blocks across the encoder and decoder portions of the network are connected by skip-like connections indicated in black arrows. Spatial uncertainty sampling via dropout is applied between successive convolutional blocks to estimate model uncertainty, and network outputs after convolutional blocks are visualized in inset.}
\label{architecture}
\end{figure*}

To perform the segmentation task, we utilized a convolutional U-Net architecture \cite{ronneberger2015u}, which uses a autoencoder structure where the input image is compressed and then decompressed through successive convolutional layers, connected via skip-like connections (Fig.~\ref{architecture}). The network was trained using the pixel-wise cross-entropy loss, 
\begin{equation}
    \mathcal{L}(y, \hat{y})= - \sum_{i} y_i \log \hat{y_i} +  (1 - y_i) \log (1 - \hat{y_i})
\end{equation}
where $i$ is the index of a pixel in an image, $y$ is the true label of a pixel, and $\hat{y} \in [0,1]$ is the predicted label of the pixel. We performed one-fold cross validation using the validation data, and applied our trained model to a held-out test set to assess performance. All models were implemented in TensorFlow and trained on a NVIDIA Titan X GPU. 

The performance of our U-Net model was benchmarked against the ground truth human annotation, a thresholding algorithm, and a regularized logistic regression model trained on class-balanced patches. We utilized a commonly used evaluation metric, the Average Precision Score (APS): $APS = \sum_n (R_n - R_{n-1}) P_n$, where $R_n$ and $P_n$ are the recall and precision, respectively, at threshold $n$. Precision is defined as the percentage of segmented pixels that correctly correspond to a parasite, and recall is the percentage of true segmented pixels identified.

\subsection{Uncertainty Estimation}
To investigate the confidence of our model in generating parasite segmentations, we utilized a dropout based approach for estimating model uncertainty \cite{gal2016dropout, gal2016theoretically, amini2018spatial}. It has been shown that applying dropout before every weight layer in a neural network is equivalent to approximating a probabilistic deep Gaussian process, and that this approximates the posterior distribution over the network weights, $q(\bm{W})$ \cite{gal2015bayesian, gal2016dropout}. Given input image data $\bm{X}$ and a segmentation output $\bm{Y}$, we can use the dropout-based approximation of the posterior, to obtain an estimation for the predictive distribution $q(\bm{Y}\vert\bm{X})$, the likelihood of a segmentation given an input image. We have:
\begin{equation}
q(\bm{Y}\vert\bm{X}) = \int P(\bm{Y}\vert \bm{X}, \bm{W}) \, q(\bm{W})\, d\bm{W}
\end{equation}
Given $T$ stochastic forward passes through the network generated with dropout, $\{ \bm{W_t} \}_{t=1}^T$, we can define the predictive mean as: $\bm{E}[\bm{Y} \vert \bm{X}] = \frac{1}{T}\sum_{t=0}^T \bm{f}(\bm{X},\bm{W_t})$. Thus, the predictive variance, which defines the model uncertainty, is:
\begin{equation}
\textrm{Var}[\bm{Y} \vert \bm{X}] = \frac{1}{T}\sum_{t=0}^{T-1} \bm{f}(\bm{X},\bm{W_t})^2 - \bm{E}[\bm{Y} \vert \bm{X}]^2.
\end{equation}

Given that the U-Net architecture is purely convolutional, we utilize spatial Bernoulli dropout \cite{tompson2015efficient, amini2018spatial} for the stochastic sampling. In this method, our uncertainty estimates result from sampling a Bernoulli random variable $\bm{Z}$ to drop entire feature maps in the network, where $\bm{z}^{(k,l)} \sim Bernoulli(p)$ corresponds to $k$-th feature map in the $l$-th layer and $p$ is the probability that all units in the feature map remain active \cite{gal2015bayesian}. This approach has previously been shown to be a special case of element-wise dropout and thus a valid way for estimating model uncertainty \cite{amini2018spatial}.

\section{Experiments and Results}

\begin{figure*}[!t]
\centering
\includegraphics[width=0.9\linewidth]{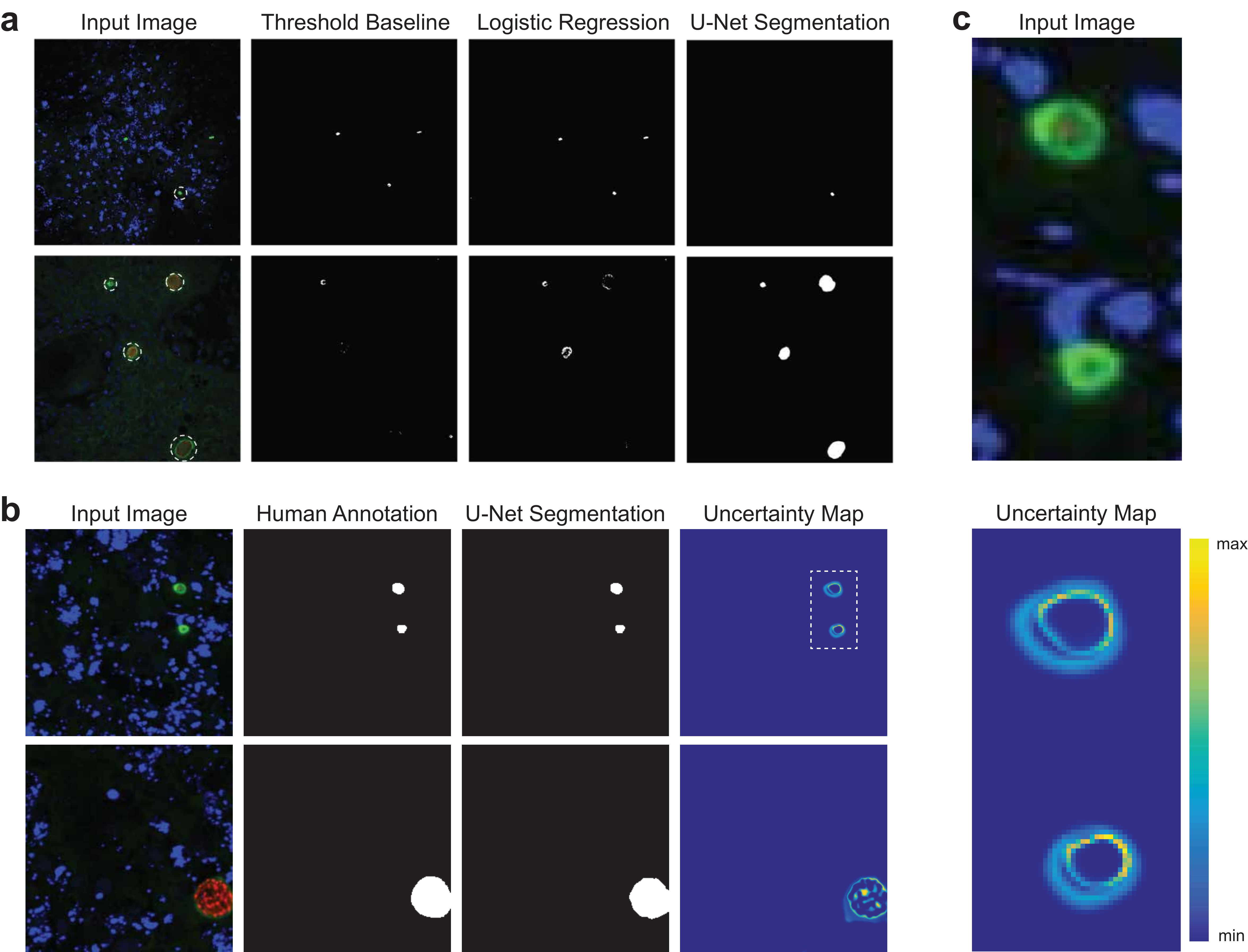}
\caption{{\bf Neural network segmentation performance and uncertainty estimation.} A: Results for the segmentation task on dataset images with true positives circled in white, comparing the segmentations returned by the threshold baseline, logistic regression, and U-Net models.
B: Representative examples from dataset (left), ground truth human annotation (middle left), U-Net segmentations (middle right), and visualization of model uncertainty (right). C: Zoom-in of the boxed region in (b) comparing the input image (top) to estimations of model uncertainty over the region of interest (bottom).}
\label{results}
\end{figure*}

To assess the performance of our model on the segmentation of \textit{P. vivax} parasites, we trained the U-Net architecture with pixel-wise cross entropy loss, evaluated our trained model on a held-out test dataset of \textit{P. vivax} liver stage culture and assessed segmentation precision using the APS. As baselines, we evaluated the performance of the thresholding and logistic regression models on the test dataset. 

Representative examples of the resulting segmentations are shown in Fig.~\ref{results}a. We visually compared the segmentations returned by each of thresholding baseline, logistic regression, and the U-Net with cross entropy loss to images from the dataset with true positives highlighted, and found that the baseline models appear unable to distinguish parasites from other stains of similar brightness, or recognize that two regions of significantly different brightnesses could both be parasites. These results suggest that our U-Net architecture is both sensitive and specific in its detection and segmentation of parasites in these images. More quantitatively, we evaluated the precision of the U-Net and logistic regression models using the APS, and found that the U-Net trained with cross-entropy loss achieved a precision exceeding 98\% on the held-out test set, while the logistic regression model achieved a precision of 80.7\%. 

In addition to the detection and segmentation of malaria parasites in hepatocyte cultures, statistics about the number and size of parasites present in an image are critical to both drug development and fundamental biology research. Using a simple clustering algorithm, we count and size parasites with high accuracy, identifying approximately 10\% for human annotation. Out of the 13.3\% of the images where the number of parasites was incorrectly counted, 91.3\% were within 1 of the true annotation.

Finally, we applied spatial dropout sampling to estimate and subsequently visualize model uncertainty, that is, the regions of predicted segmentation where we can think of the model as being less confident in its prediction (Fig.~\ref{results}b, c). As shown in Fig.~\ref{results}c, the U-Net model is more uncertain in its prediction around the edges and membranes of the segmented parasites, and more certain in the interior, cytoplasmic portion of the parasites, consistent with what we may expect in terms of a model's confidence in detection and segmentation of particular regions of an image. It is important to note that while the network's predicted \textit{probability} for a given class may be high, the model may not necessarily be \textit{confident} in that prediction. The uncertainty algorithm presented here may thus be used to capture these situations and to inform subsequent human annotation.

\section{Conclusion}

Our work supports the potential for neural network-based methods to automate previously time-intensive processes of the anti-malarial drug development pipeline. In particular, our U-Net architecture demonstrates robust segmentation of parasites during the liver stage of \textit{P. vivax} malaria with near-human accuracy. Furthermore, we develop a novel method for uncertainty estimation in image segmentation tasks, utilizing spatial dropout sampling to visualize model uncertainty and evaluating our approach on the malaria dataset. We believe that our neural network approach provides an efficient, accurate method for automated image segmentation of liver stage malaria for applications in drug development and screening, and that our spatial dropout algorithm provides a generalizable method for robust estimation of model uncertainty in image segmentation tasks.

\newpage
\bibliography{icml_malaria}
\bibliographystyle{icml2019}

\end{document}